\documentclass[aps,pre,reprint,onecolumn,notitlepage,groupedaddress]{revtex4-1}
\usepackage{graphics,amssymb,amsmath,color,bm,mathtools,upgreek,graphicx,float,xcolor,cancel,mathtools}
\usepackage[colorlinks=true,linkcolor=blue]{hyperref}
\usepackage[sfdefault=cmbr]{isomath}
\usepackage[toc,page]{appendix}

\numberwithin{equation}{section}

\definecolor{navyblue}{RGB}{0,0,128}
\definecolor{dodgerblue}{RGB}{30,144,255}
\definecolor{darkgrey}{RGB}{169,169,169}
\definecolor{deepskyblue}{RGB}{0, 191, 255}

\usepackage{bm}
\def\boldsymbol{\bm}

\def \t{\tensorsym}
\def \lb{\left}
\def \rb{\right}

\def \d{\,\text{d}}

\def \etah{\hat{\eta}}

\def \bgamma{\boldsymbol{\gamma}}

\def \bgammad{\dot{\boldsymbol{\gamma}}}
\def \bgammadh{\hat{{\dot{\bgamma}}}}
\def \gammad{{\dot{\gamma}}}

\def \bOmega{\bm{\Omega}}

\def \bsigma{\boldsymbol{\sigma}}

\def \bsigmah{\hat{\boldsymbol{\sigma}}}

\def \btau{\boldsymbol{\tau}}

\def \bTheta{\bm{\Theta}}

\def \bzero{\bm{0}}

\def \fB{\mathcal{B}}

\def \tEh{\mathsf{\t{\hat{E}}}}

\def \bF{\bm{F}}

\def \tF{\mathsf{\t F}}
\def \tFh{\mathsf{\t{\hat{F}}}}

\def \bI{\bm{I}}

\def \fO{\mathcal{O}}

\def \bR{\bm{R}}
\def \bRh{\hat{\bm{R}}}

\def \tRh{\mathsf{\t{\hat{R}}}}

\def \bS{\bm{S}}

\def \tTh{\mathsf{\t{\hat{T}}}}

\def \bU{\bm{U}}

\def \tU{\mathsf{\t U}}
\def \tUh{\mathsf{\t{\hat{U}}}}

\def \fV{\mathcal{V}}

\def \be{\bm{e}}

\def \bn{\bm{n}}

\def \br{\bm{r}}

\def \bu{\bm{u}}

\def \bx{\bm{x}}

\usepackage{scalerel,stackengine}
\stackMath
\newcommand\reallywidehat[1]{%
\savestack{\tmpbox}{\stretchto{%
  \scaleto{%
    \scalerel*[\widthof{\ensuremath{#1}}]{\kern-.6pt\bigwedge\kern-.6pt}%
    {\rule[-\textheight/2]{1ex}{\textheight}}
  }{\textheight}%
}{0.5ex}}%
\stackon[1pt]{#1}{\tmpbox}%
}
\begin{document}
\title{Dynamics and rheology of particles in shear-thinning fluids}
\author{Charu Datt}
\affiliation{Department of Mechanical Engineering, 
University of British Columbia,
Vancouver, BC, V6T 1Z4, Canada
}

\author{Gwynn J. Elfring}\email{Electronic mail: gelfring@mech.ubc.ca}
\affiliation{Department of Mechanical Engineering, 
University of British Columbia,
Vancouver, BC, V6T 1Z4, Canada
}
\date{\today}

\begin{abstract}
Particle motion in non-Newtonian fluids can be markedly different than in Newtonian fluids. Here we look at the change in dynamics for a few problems involving rigid spherical particles in shear-thinning fluids in the absence of inertia. We give analytical formulas for sedimenting spheres, obtained by means of the reciprocal theorem, and demonstrate quantitively differences in comparison to a Newtonian fluid. We also calculate the first correction to the suspension viscosity, the Einstein viscosity, for a dilute suspension of spheres in a weakly shear-thinning fluid. 
\end{abstract}

\maketitle

\section{Introduction}
\label{intro}

Particles in fluids are ubiquitous in both natural and industrial processes.  Blood, detergents, paints, aerated drinks, fibre-reinforced polymers, sewage sludges, and drilling muds are some examples where particles ---rigid or drops or bubbles--- are present in a suspending fluid \citep{barnes_rheology_review, chhabra2006bubbles}. The flow behaviour and rheological properties of such suspensions depend on parameters like the particles' shape, size and concentration, particle-particle interaction, particle surface properties, fluid rheology and the type of flow.  Even the simplest of such suspensions -- small, rigid non-Brownian particles in a Newtonian fluid -- exhibits rich rheological properties like shear-thinning, shear-thickening, and normal stress differences which are characteristics of complex fluids \citep{Morris_review, powell_review}. In many common examples like paints, foods, fracking fluids and biological suspensions, the suspending fluid itself is non-Newtonian. The properties of these suspensions is therefore expected to be even more \textit{complex} \citep{ landel_susp_thin, shaqfeh_susp_elastic, Metzner_1985}.  

In order to understand the properties of these suspensions, it becomes imperative to first understand the interaction of a single particle with the surrounding media and the mutual interaction between two such particles. In fact, it is known that the motion and orientational dynamics of particles can be strongly affected by the rheology of the surrounding fluid medium \citep{brunn_review, leal79, leal80, D_Avino_2015, roberto_feng}. For example, at zero Reynolds number, while the center-to-center distance between two equal spherical particles which are sedimenting along their line of centers through a quiescent fluid is fixed indefinitely at its initial value in a Newtonian fluid \citep{stimson1926motion}, this distance is found to change in the presence of viscoelasticity of the fluid medium \citep{RIDDLE197723}.   The understanding of particle dynamics in complex fluids is also important for applications in particle manipulation in microfluidic devices (see recent reviews \citep{Lu_2017, D_Avino_2017}).  Phenomena like cross-stream migration, in which rigid spheres in a pressure-driven tube flow of viscoelastic liquid migrate either towards or away from the wall in the absence of inertia, can be used for cell-trapping in biomedical applications \citep{Karimi_2013}. 

Towards a fundamental understanding of particle dynamics in non-Newtonian fluids, in this work, we theoretically study the dynamics of rigid non-Browinian spherical particles in shear-thinning fluids in the absence of any fluid or particle inertia. Unlike for viscoelastic fluids, where theoretical studies have been used to develop insights for many experimental observations \citep{ leal79,D_Avino_2015}, similar studies have been relatively few in shear-thinning fluids and most of these studies have focussed on using the power-law model \citep{bird1987dynamics} to model the shear-thinning rheology \cite{chhabra2006bubbles}. However, as argued by \citet{Chhabra1980}, a fluid model with a zero-shear viscosity should be preferred to the power-law model for slow flows around spheres. Here we use the Carreau model for shear-thinning fluids \citep{bird1987dynamics} (discussed in the subsequent section) to study the following problems motivated by some recent experiments:

i) Two equal spherical particles sedimenting along their line of centres through a quiescent fluid. In Newtonian fluids, \citet{stimson1926motion} showed that the initial distance of separation is maintained as the particles sediment. 

ii) Sedimentation of a spherical particle which is also rotating due to some external field. In Newtonian fluids, the sedimenting velocity does not depend on the rotation rate. The translational and rotational motion for a sphere are decoupled in a Newtonian fluid \citep{happel2012low}. 

iii) Sedimentation of a spherical particle in a linear background flow. In Newtonian fluids, the sedimenting velocity of a sphere depends only on the (local) velocity of the background flow but is independent of the velocity gradient.

iv) The influence of particles on the viscosity of a shear-thinning fluid. For a dilute suspension of neutrally buoyant particles in a Newtonian fluid, it was shown by Einstein \citep{einstein1956investigations} that the bulk shear viscosity of the suspension increases due to the presence of particles.

In the following sections, we analyse these problems in shear-thinning fluids in detail, but before that we briefly discuss our theoretical approach and the rheology of shear-thinning fluids.

\section{Reciprocal Theorem}
We are interested here in the motion of, or equivalently forces on, particles in complex fluids. These integrated quantities can be evaluated without resolution of the associated flow field of the complex fluid by employing the reciprocal theorem. This approach was comprehensively reviewed by \citet{leal80}, and we use here a generalized formalism developed in a number of recent papers for active particles \citep{datt17, elfring15, elfring16, elfring17}. Following \citet{elfring17}, the motion $\tU$ or forces $\tF$ of $N$ particles in a complex flow may be given by
\begin{align}\label{reciprocalmain}
\tU=\frac{\etah}{\eta}\tRh_{\tF\tU}^{-1}\cdot\lb[-\tF+\tF_T+\tF_{NN}\rb],
\end{align}
where $\tU=[\bm{U} \ \bm{\Omega}]$, $\tF=[\bm{F} \ \bm{L}]$ are $6N$-dimensional vectors comprising translation/rotation and hydrodynamic force/torque respectively on $N$ particles. If the inertia of the particles is negligible (small Stokes numbers), as we will assume here, then the hydrodynamic force must balance any external or applied force (such as weight due to gravity)  $\tF = - \tF_{ext}$. The force
\begin{align}
\tF_T = \frac{\eta}{\etah}\int_{\partial\fB}\lb(\bu^S-\bu^\infty\rb)\cdot(\bn\cdot\tTh_\tU)\d S,
\end{align}
is a Newtonian `thrust' due to any surface deformation or activity of the particles $\bu^S$ (although in all cases here we consider rigid passive particles, $\bu^S=\bzero$) and `drag' from any background flow $\bu^\infty$. Here $\partial\fB$ represents the surfaces of all the particles. The non-Newtonian contribution
\begin{align}\label{forcenn}
\tF_{NN} = -\int_{\fV}\btau'_{NN}:\tEh_{\tU}\d V,
\end{align}
represents the extra force/torque on each particle due to a non-Newtonian deviatoric disturbance stress $\btau'_{NN}=\btau_{NN}-\btau^{\infty}_{NN}$ in the fluid volume $\fV$ in which the particles are immersed.

The formulas rely on operators from an $N$-body resistance/mobility problem in a Newtonian fluid (with viscosity $\hat{\eta}$)
\begin{align}
\bgammadh' / 2 &= \tEh_{\tU}\cdot\tUh',\label{linearE}\\
\bsigmah' &= \tTh_{\tU}\cdot\tUh',\label{linearT}\\
\tFh' &= -\tRh_{\tF\tU}\cdot\tUh'\label{linearR}.
\end{align}
where primes indicate disturbance quantities. The tensors $\tEh$ and $\tTh$ are functions of position in space that map the (arbitrary) motion of all $N$ particles $\tUh$ to the fluid strain-rate and stress fields respectively, while the $N$-body rigid-body resistance tensor
\begin{align}
\tRh_{\tF\tU} = 
\begin{bmatrix}
\bRh_{FU} & \bRh_{F\Omega}\\
\bRh_{LU} & \bRh_{L\Omega}
\end{bmatrix}.
\end{align} 
We note that no specific $\tUh$ needs to be chosen in the rigid-body dual problem as only the linear operators $\tEh$, $\tTh$ and $\tRh$ enter the picture. In many cases there may be symmetries in the problem which simplify these operators substantially, likewise we may know that the forces/torques are in some way simplified (collinear with gravity for instance) and hence need not even determine all components of the operators.

\section {Shear-thinning fluid}
\label{sec_non_dim}
As we outlined in the previous sections, the presence of a non-Newtonian stress $\btau_{NN}$ can significantly alter the motion of particles in flows. In this paper, we consider the effects of shear-thinning fluids, which experience a loss in apparent viscosity $\eta$ with increasing strain-rates $\gammad$; specifically, the deviatoric stress
\begin{align}
\btau=\eta(\bgammad)\bgammad,
\end{align}
where the viscosity is modelled using the Carreau model for generalised Newtonian fluids \citep{bird1987dynamics} 
\begin{align}
\eta \left( \dot{\gamma}\right) = \eta_{\infty} + \left( \eta_0 - \eta_{\infty}\right) \left[ 1+ \lambda_t^2 \vert \dot{\gamma}\vert^2 \right]^{(n-1)/2}.
\label{consti_dim}
\end{align}
Here, $\eta_0$ is the zero-shear-rate viscosity, $\eta_{\infty}$ is the infinite-shear rate viscosity, $n$ is the power law index ($ n< 1$ for shear-thinning fluids; smaller the value of $n$, more shear-thinning the fluid is) and $\lambda_t$ is a time constant.   The magnitude of the strain-rate is given by $\vert \dot{\gamma}\vert = \left( \Pi\right)^{1/2}$ where $\Pi = \dot{\gamma}_{ij} \dot{\gamma}_{ij}$ is the second invariant of the strain-rate tensor.  Note that $\lim_{\dot{\gamma} \to 0}\eta \left( \dot {\gamma} \right) = \eta_0$ and $\lim_{\dot{\gamma} \to \infty}\eta \left( \dot {\gamma} \right) = \eta_{\infty}$, showing that at low and high strain rates the fluid behaves like a Newtonian fluid with viscosity $\eta_0$ and $\eta_{\infty}$, respectively. $\lambda_t$ sets the cross-over strain-rates at which non-Newtonian effects start to become important. 

In this work, we investigate strain-rates such that $\lambda_t \ll 1/ \dot{\gamma_c}$, where $\dot{\gamma}_c$ is the characteristic strain-rate of the flow. In this case it is useful to write the constitutive equation in the form
\begin{align}
\btau=\eta_0\bgammad+ \left(\eta(\bgammad)-\eta_0\right)\bgammad.
\end{align}
Although, we note that this rearrangement is not in any way restricted to low strain rates. Writing the equation as such, it is clear that the non-Newtonian contribution $\btau_{NN}= (\eta(\bgammad)-\eta_0)\bgammad$.

In dimensionless form,  one may decouple the Newtonian and non-Newtonian contribution for a Carreau fluid as
 \begin{align}
\boldsymbol{\tau}^* = \bgammad^*+\left\{\beta-1 + \left( 1- \beta\right) \left[ 1+ Cu^2 \vert \dot{\gamma}^* \vert ^2 \right]^{{\left(n-1\right)}/{2}}\right\} \boldsymbol{\dot{\gamma}}^*,
\label{consti_eq}
\end{align}
where stars (*) represent dimensionless flow quantities. The Carreau number $Cu = \dot{\gamma_c} \lambda_t$ is the ratio of the characteristic strain rate in the flow $\dot{\gamma_c}$ to the crossover strain rate $1/ \lambda_t$. The viscosity ratio is given by $\beta= \eta_{\infty} / \eta_0 \in [0,1]$. The characteristic length of the particle is chosen as the length scale in the problems and as we consider only spherical particles, the length scale is $a$, the radius of the particles. $\eta_0 \dot{\gamma_c}$ is the scale for stresses; the appropriate characteristic strain-rate, $\dot{\gamma}_c$, varies depending on the problem and therefore, is defined separately in each of the problems below.

In this work, we consider the fluid behaviour to be weakly shear-thinning, in the sense that $Cu\ll1$ \citep{jfmCharu}, and therefore the viscosity is assumed to not deviate substantially from the zero-shear viscosity $\eta_0$. We then explore the leading-order weakly shear-thinning effects of the fluid rheology on particle motion. To this end, we assume a regular perturbation expansion of all fields, e.g. $\bm{u} = \bm{u}_0+Cu^2\bm{u}_1+\ldots$, and find the non-Newtonian stress to be
\begin{align}\label{taunn}
\boldsymbol{\tau}^*_{NN} = -\frac{1}{2}Cu^2\left(1- n\right) \left( 1 - \beta \right) \vert \dot{\gamma}^*_0 \vert^2 \dot{\boldsymbol{\gamma}}^*_0+\fO(Cu^4),
\end{align}
where for shear-thinning fluids $\beta < 1$ and $n<1$.  We consider only the leading-order effects of shear-thinning viscosity and by using \eqref{taunn}, we may obtain the non-Newtonian force on particles at the expense of an integration  \eqref{forcenn} which then only requires the Newtonian flow field $\bm{u}_0$.

\section{Sedimenting spheres}
Due to their symmetry, a number of classic results involving motion of spheres in Newtonian fluids can be predicted directly by employing the kinematic reversibility of the field equations and it is insightful to consider examples where dynamics are altered (or not) by shear-thinning rheology.  In this section, we explore one such case of two spheres sedimenting along their line of centres but before that, it is instructive to first examine the simple case  of a single sphere moving through a shear-thinning fluid and to see how the force-motion relationship is affected by the medium rheology. 

\subsection{Single sphere}
\label{single_sphere}
 The drag force on a sphere of radius $a$, moving with a velocity $\bm{U}$ is given by $\bm{F} = - 6 \pi a \eta \bm{U}$ (dimensional), where $\eta$ is the viscosity of the fluid. In a shear-thinning fluid with zero-shear rate viscosity $\eta_0 =\eta$, the drag force is expected to be less than the Newtonian value. This is because one expects the apparent viscosity around the sphere to decrease below $\eta$ due to the strain-rates ensuing from the motion of the sphere. Quantitatively, the drag force in shear-thinning fluid can be evaluated using the reciprocal theorem. For a single sphere, we expect this force to be colinear with the velocity by symmetry. Simplifying \eqref{forcenn} we may write
\begin{align}
\bm{F} = -6\pi \eta_0 a \bm{U} - \int_{\fV} \btau_{NN} : \tEh_U \d V, 
\label{final_eq}
\end{align}
where $\tEh_U$ (such that $\bgammadh/2 = \tEh_U\cdot\hat{\bm{U}}$) is well known for a sphere translating in Stokes flow \citep{Guazzelli_2009}. The integral above may be easily evaluated to leading order as then the non-Newtonian stress depends only on the solution of a translating sphere in a Newtonian fluid i.e. $\btau_{NN}[\bm{u}_0]$ (from\eqref{taunn}). In dimensionless form we find the drag in a shear-thinning fluid to be
\begin{align}
\bm{F}^{*} = - 6 \pi \bU^*\left( 1  - \frac{1}{2}\left(1 - \beta\right) \left(1 -n \right) Cu^2 \dfrac{942}{2275}\lb|\bU^*\rb|^2 \right).
\label{single_sphere_eq}
\end{align}
If we (sensibly) take as the characteristic strain-rate $\gammad_c=\vert \bU \vert/a$ then $\bU^*=\be$ is simply the unit vector in the direction of the motion (a convention we use below). The term in the brackets is then an analytical calculation of the drag correction factor (often given the symbol $X$ in the literature \citep{Chhabra1980a, BUSH_Phan}) valid to $\fO(Cu^2)$.

We see that, as expected, the drag force on a sphere decreases in a shear-thinning fluid as compared to a Newtonian fluid. It should be noted that the reduction in drag below the Newtonian value is at odds with the conclusions reached using the power-law fluid model which predict an increase in the drag force \citep{chhabra2006bubbles} because (as argued in that work) the power-law fluid model does not incorporate a zero shear-rate viscosity which is important in modelling slow flows involving stagnation points and vanishingly small shear rates.  In contrast, \eqref{single_sphere_eq} predicts the reduction in drag observed in experimental results \citep{Chhabra1980a}, and qualitatively agrees with a variational estimate by \citet{Chhabra1980a}, and numerical results from \citet{BUSH_Phan}, which both used the Carreau fluid model to characterise the fluid rheology.

It is straightforward to invert the drag force to obtain the velocity given a prescribed external force (for example weight due to gravity in sedimentation)
\begin{align}
\bU^{*} = \frac{\bF^*_{ext}}{6\pi}\lb(1 +\frac{1}{2}\left(1 - \beta\right) \left(1 -n \right) Cu^2 \dfrac{942}{2275}\frac{\lb|\bF_{ext}^*\rb|^2}{(6\pi)^2}\rb).
\label{mobility}
\end{align}
In this case an appropriate strain-rate scale is $\gammad_c = \vert \bm{F}_{ext} \vert /  \left(\eta a^2 \right) $ in which case $\bF^*_{ext}$ would be a unit vector.  

\subsection{Two spheres} 

We now consider sedimentation of two spheres of equal radii along the line joining their centres. In a Newtonian fluid, one can use arguments of kinematic reversibility and symmetry to find that the two spheres will sediment with equal velocities and will maintain their initial distance of separation \citep{Guazzelli_2009}. \citet{stimson1926motion} solved the hydrodynamically equivalent problem of two spheres moving with a constant velocity along their line of centres and calculated the flow field and the forces on the spheres. When their radii are equal,  it was found that the forces on each of the two spheres are indeed equal but each less than on single sphere moving in a quiescent fluid with the same velocity.  Quantitatively, the force on either sphere can be written as $\bm{F} = -6\pi \eta a \bm{U} \lambda$ (dimensional), where $\lambda$ is a coefficient which depends on the separation between the two spheres \citep{stimson1926motion}.  $\lambda \to 1$ as the distance between the two spheres approaches infinity, i.e. when the two spheres do not interact hydrodynamically with each other. It was also found that value of $\lambda$ decreases as the distance between the two spheres decreases. However, the method of \citet{stimson1926motion} could not be applied to the case of two spheres in contact with each other.  This case was later solved by \citet{cooley_1969} who found $\lambda = 0.645$ for the two sphere touching case. These results and values of $\lambda$ have also been observed experimentally \citep{happel2012low}.

Non-Newtonian rheology can break kinematic reversibility and indeed viscoelasticity is known to change these results qualitatively. Where in Newtonian fluids two equal sedimenting spheres maintain their initial distance of separation, it was  found in the experiments of \citet{RIDDLE197723} and analyses of \citet{ Brunn1977} and \citet{two_sphere_viscoelastic} that the two particles showed a tendency to aggregate in viscoelastic fluids. This effect of normal stresses on particle dynamics has been commented upon by \citet{jimmy_second_order}.

In shear-thinning fluids, it was observed in the experiments of \citet{daugan2002aggregation} that the two spheres would aggregate provided the initial distance of their separation was smaller than some critical distance. The numerical study by  \citet{yu2006numerical}, the authors argued that in fact this tendency towards aggregation was due to thixotropy (memory of shear-thinning) and the corridors of reduced viscosity in the wake of sedimenting particles lead to aggregation \citep{joseph_94,daugan2002aggregation}. In the absence of memory, the two spheres would maintain their initial distance of separation \citep{yu2006numerical}. Here, we theoretically study the equivalent problem of two equal spheres moving with a constant velocity along their line of centres, as considered by \citet{stimson1926motion}, but in a shear-thinning fluid. 

We use the reciprocal theorem to calculate the forces on the two spheres. By \eqref{reciprocalmain} the force on the particles can be written generally as
\begin{align}
\tF=-\frac{\eta}{\etah}\tRh_{\tF\tU}\cdot\tU+\tF_{NN}.
\end{align}
For two equal spheres moving along their line of centres, by symmetry, we expect all vectors in the problem to be collinear, which significantly simplifies the more general problem of the motion of two spheres. When the motion of two bodies is collinear (only translational motion is considered), it is useful to decompose the motion into a mean velocity $\overline{\bm{U}}$ and relative velocity $\Delta\bm{U}$ such that the velocity of each sphere may be written as $\bm{U}_1 = \overline{\bm{U}}+ \Delta\bm{U}$ and $\bm{U}_2 = \overline{\bm{U}}-\Delta\bm{U}$. In this basis, the relevant resistance/mobility problems for the reciprocal theorem are i) two (equal) spheres translating with equal velocity along the line joining their centres in a Newtonian fluid (corresponding to  $\overline{\bm{U}}$ as in this case $\Delta\bm{U} = 0$) with solution by \citet{stimson1926motion} and ii) two (equal) spheres approaching each other with equal speed along the line joining their centres in a Newtonian fluid with solution due to \citet{maude1961end} and \citet{BRENNER1961242} (corresponding to  $\Delta\bm{U} $ as in this case $\overline{\bm{U}} = 0$). Note that one requires two resistance/mobility problems for evaluating the force on each of the two spheres.  The resistance tensor for this problem is diagonal in the sense that in a Newtonian fluid mean translation leads only to a mean force and likewise for relative force. The reciprocal theorem given above leads to expression for mean and relative force in a non-Newtowian fluid given by

\begin{align}
\overline{\bm{F}}   &= -\frac{\eta_0}{\etah}\hat{\bR}_{\overline{F}\overline{U}} \cdot \overline{\bm{U}} - \frac{1}{2}\int_{\fV} \btau_{NN} : \tEh_{\overline{U}} \d V,
\label{solve_1}\\
\Delta\bm{F} &= -\frac{1}{2}\int_{\fV} \btau_{NN} : \tEh_{\Delta U} \d V,
\label{solve_2}
\end{align}
where $\hat{R}_{\overline{F}\overline{U}}$ is the (mean) translational resistance of the two spheres in a Newtonian fluid and $\tEh_{\overline{U}}$ and $\tEh_{\Delta U}$ correspond to the strain-rate due to mean and relative motion respectively. Note that $\btau_{NN}$ is evaluated using the solution of the problem in Newtonian fluids by \citet{stimson1926motion} (from \eqref{taunn}).

Upon evaluation of \eqref{solve_2}, for weakly non-linear shear-thinning fluids one finds that there is no relative force
\begin{align}
\Delta\bF=\bzero,
\end{align}
meaning the forces on two equal spheres are equal in a shear-thinning fluid as in a Newtonian fluid. Although we obtain this result only for a weakly shear-thinning fluid, we expect this to be the case for all generalized Newtonian fluids, regardless of the parameter regime. The reason is that the stress $\btau_{NN}$ maintains the symmetry of the Newtonian problem while the Maude-Brenner problem (and thus the operator $\tEh_{\Delta U}$) displays a mirror-image symmetry and therefore the integral over the entire fluid volume must be zero. In fact, \citet{brunn_review} briefly comments on this property of generalized Newtonian fluids where results may come out to be similar to Newtonian fluids.

\begin{figure}
\centering
\includegraphics[width = 0.95\textwidth]{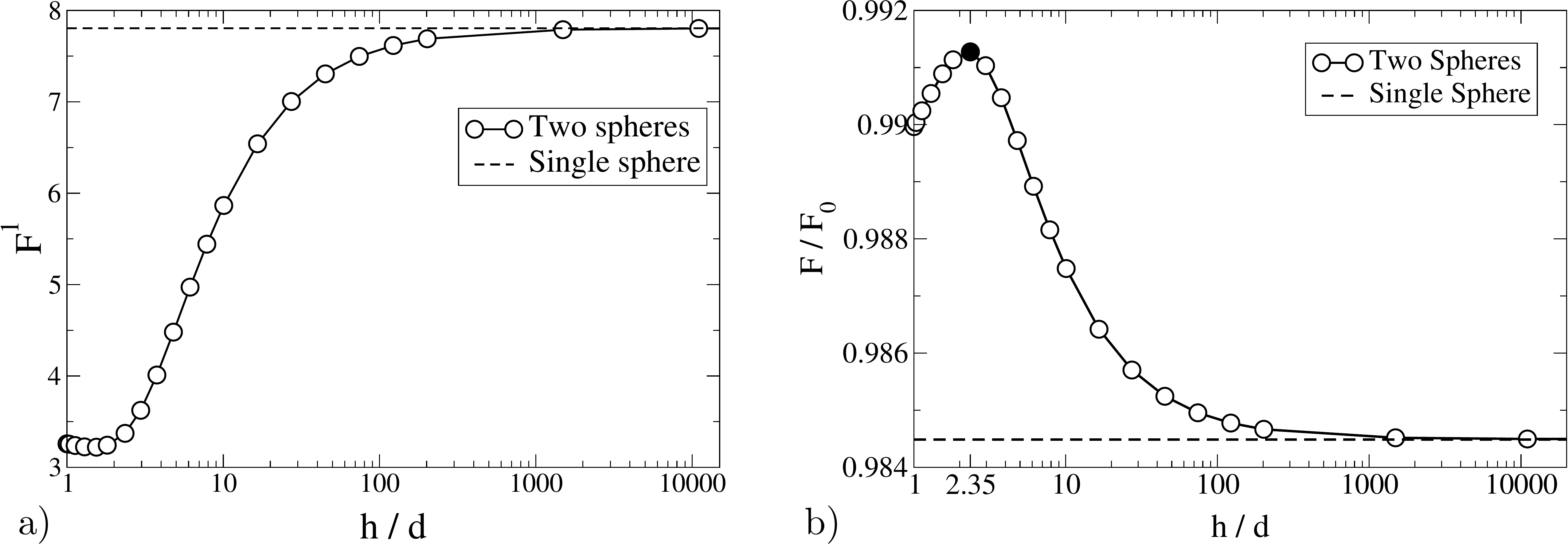}
\caption{a) Variation of non-Newtonian drag $F^1$ as a function of the distance of separation between the two spheres.\\ b) Normalized drag as a function of this distance. Note that the variation is non-monotonic and that drag reduction is minimized at $h/d \approx 2.35$.}
\label{fig1}
\end{figure}

As the force on the two spheres in a weakly shear-thinning fluid are equal, two sedimenting spheres do not show any tendency to aggregate in a shear-thinning fluid without memory, as was also found in the numerical work of \citet{yu2006numerical} discussed previously. We can further calculate the force on each of the spheres, and compare it to the force on a single sphere in a shear-thinning fluid from section \ref{single_sphere}. Since there is no difference in drag force, the mean in \eqref{solve_1} represents the hydrodynamic drag on each sphere. The form of the force (dimensionless) is
\begin{align}
\overline{\bF}^*   &=  \overline{\bF}^*_0 + \frac{1}{2}Cu^2\left(1- \beta\right)\left(1- n \right) F^{1}\be+\fO(Cu^4),
\end{align}
where $\overline{\bF}^*_0$ is the force on each sphere in a Newtonian fluid \citep{stimson1926motion}. Here we have non-dimensionalised lengths with sphere radius $a$ and stresses with $\eta_0 U/ a$ where $U$ is the magnitude of velocity of the spheres such that $\bU^*=\be$.  Both $\overline{\bF}^*_0$ and the coefficient $F^1$ obtained by evaluating the integral in \eqref{solve_1} numerically, depend on the ratio of the centre distance between the spheres to their diameter, $h/d$, as shown in Figure \ref{fig1}a). Note that $F^1$ is positive, meaning the correction to drag force is in the direction of the motion $\be$ and so the drag in a shear-thinning fluid is less than in a Newtonian fluid. To contrast the results with the Newtonian fluids case, we also plot  the ratio of the magnitude of the force in a weakly shear-thinning fluid (correct to $\fO(Cu^2)$) to the drag in an equivalent Newtonian fluid for same configuration in Figure \ref{fig1}b).  This is plotted in Figure \ref{fig1}b) for $Cu^2 = 0.1$, $\beta = 0.001$ and $n = 0.25$. 

 We note that the ratio $F/ F_0$ is always less than 1 which shows that the drag in a shear-thinning fluid is less than that in Newtonian fluid for the same configuration. This is expected as the viscosity in a shear-thinning fluid decreases with strain-rate leading to less drag on each sphere when compared to in Newtonian fluid. But what is perhaps surprising is the variation of force with distance between the two spheres. We first note that as this distance becomes large, the drag reduction on each sphere asymptotes to that on a single sphere in shear-thinning fluid. The drag reduction when the two spheres essentially do not interact hydrodynamically with each other is greater than for any other distance of separation.  Shear-thinning effects are maximum in this configuration. When the spheres interact hydrodynamically the effective strain rates are reduced (due to screening) and thus the drag reduction on each is always lower than for a single sphere. On the other end of the distances, we have the configuration when the two spheres are in contact with each other.  This, interestingly, is not the configuration to observe minimum shear-thinning effects. In fact, the minimum shear-thinning effects or equivalently the maximum of the ratio $F/ F_0$ is observed when the clearance between the two spheres is of the order of their diameter i.e. when $h/d \approx 2.35$. We believe that this nontrivial result may be due to the complex flow field around the spheres, which also includes ring vortices, and its effect on the dissipation rates \cite{O_neill_1976}.

\section{Sedimentation of a rotating sphere}
We now consider the case of a sphere which translates as well as rotates in a shear-thinning fluid. The calculations are inspired by the recent experimental work of \citet{Godinez2014} who study the hydrodynamically equivalent problem of sedimentation of a rotating sphere in a power-law fluid. By imposing a controlled rotation on a sedimenting sphere, \citet{Godinez2014} measured the increase in the sedimentation velocity, which could then be used to predict the values of power law indices of the fluids. They considered rotation of the sphere only about the sedimenting axis. Here we consider the problem more generally. 

Translation and rotation of a sphere in a Newtonian fluid are decoupled and owing to the linearity of the Stokes flow, one may superimpose the solution of translation alone and rotation alone to get the solution of a translating-rotating sphere in a Newtonian fluid. In other words a sphere that rotates in a Newtonian fluid will sediment at the same rate as when it does not rotate. This decoupling of translation and rotation is not expected to hold in a non-linear fluid. We explore this for a weakly shear-thinning fluid, again using the reciprocal theorem. 

According to the reciprocal theorem, as before, we have
\begin{align}
\tF=-\frac{\eta}{\etah}\tRh_{\tF\tU}\cdot\tU - \int_{\fV}\btau_{NN}:\tEh_{\tU}\d V.
\label{expression_for_force}
\end{align}
We non-dimensionalise length with the sphere radius, $a$, stresses by $\eta_0 U/a$ where $U$ is the magnitude of velocity of the sphere and hence $\bU^*=\be$. We consider a general angular velocity which in dimensionless form, $\boldsymbol{\Omega}^*$, is not necessarily a unit vector. Then from \eqref{expression_for_force} we get 
\begin{align}
\bm{F}^ * &= -6 \pi \be\cdot\lb\{\bI-\frac{1}{2}Cu^2\left( 1- n\right) \left( 1- \beta \right)\lb(\frac{942}{2275}\bI+\frac{552}{385}\lb[2\lb|\bOmega^*\rb|^2\bI -\bOmega^*\bOmega^*\rb]\rb)\rb\}.
\label{force_rot}
\end{align}
In the absence of any rotation $\boldsymbol{\Omega}^*=\bzero$, the force corresponds with \eqref{single_sphere_eq} as expected. With nonzero rotation, the drag force is further reduced due to the additional strain-rate caused by rotation. When the rotation is aligned with the translation, $\bOmega\propto \bU$, for example when the axis of rotation is aligned with gravity for a sedimenting sphere as in the experiments of \citet{Godinez2014}, the drag force remains collinear with $\bU$.  When the rotation is not aligned with translation, a lateral force may arise due to the term in the direction of the axis of rotation $\propto (\be\cdot\bOmega^*)\bOmega^*$. When the rotation is orthogonal to translation there is no lateral drift and the change in the drag force due to rotation is twice that of when the rotation is aligned with translation and so would maximize sedimentation velocity for a given rotation rate. Conversely, in the mobility problem for a given external force $\bm{F}_{ext}^*$, a rotating sphere will sediment with a translational velocity given by
\begin{equation}
\bm{U}^* = \dfrac{\bm{F}^*_{ext}}{6 \pi} \cdot \lb\{ \bI + \frac{1}{2}Cu^2\left( 1- n\right) \left( 1- \beta \right) \left( \dfrac{942}{2275} \dfrac{\vert \bm{F_{ext}}^* \vert ^2}{\left( 6 \pi\right)^2} \bI  + \dfrac{552}{385} \left[  2\lb|\bOmega^*\rb|^2\bI -\bOmega^*\bOmega^*\right] \right) \rb\},
\label{mobility_rotation}
\end{equation}
where $\bm{F}^*_{ext}$ is a unit vector if the strain rate scale $\dot{\gamma}_c$ is chosen as $ \vert \bm{F}_{ext}\vert / \left(\eta a^2 \right) $. On comparison with the experimental results, it is noted that \citet{Godinez2014} find a power law dependence of the sedimenting velocity on the rotation rate, $\vert \bm{U} \vert \propto \vert \bm{\Omega} \vert ^ {(1- n )}$, across a range of rotation rates such that strain-rate around the sphere is predominantly due to rotation and not due to translation, our analytical result in equation \eqref{mobility_rotation}, valid for small Carreau numbers, draws a similar picture, namely, increasing the rotation rate and decreasing the power-law index, $n$, increase the sedimentation velocity of the sphere.

We also calculate the hydrodynamic torque on the particle 
\begin{align}
\bm{L}^* = - 8 \pi \boldsymbol{\Omega}^*\cdot\lb\{\bI -  \frac{1}{2}Cu^2\left( 1- n\right) \left( 1- \beta \right)\left(\frac{24}{5}\lb|\bOmega^*\rb|^2\bI+\frac{414}{385}\lb[2\bI-\be\be\rb]  \right)\rb\}.    
\end{align}
The first term in the shear-thinning correction is due to the particle rotation alone as there is a reduction in the torque due to the shear-thinning caused by the rotation. The second term in the correction is due to both the translation and rotation of the sphere and may generate a torque which is not aligned with the direction of rotation. Clearly, in a shear-thinning fluid translational and rotational dynamics are coupled.

\section{Sphere under an external force in a linear flow of shear-thinning fluid}

We now consider the dynamics of a spherical particle driven by an external force in an unbounded linear-flow. In a Newtonian fluid, we know that the sedimenting velocity of a sphere is not altered by the velocity gradient in simple shear flow. However, this may not be the case in non-linear fluids.   In fact, in viscoelastic fluids, it is found that the terminal velocity of a sphere decreases when  the applied shear-flow is perpendicular to gravity in what is called a cross-shear-flow  \citep{davino, tanner, vandenbrule}. \citet{Gheissary_brule} used sedimentation of a sphere in cross-shear flow to predict the rheological properties of different shear-thinning fluids. The cross-shear flow is a model system used for transport of particles in hydraulically-induced fractures \citep{Mckinley_review} . \citet{mehlig_einar} recently extended the analyses in viscoelastic fluids to the case when gravity (or another external force) and the vorticity direction of the applied flow are not aligned. Here, we perform a similar analysis for a shear-thinning fluid.

Using the reciprocal theorem we calculate the velocities (both rotational and angular) of the particle as 
\begin{align}
\tU=\frac{\etah}{\eta}\tRh_{\tF\tU}^{-1}\cdot\lb[\tF_{ext}-\frac{\eta}{\etah}\int_{\partial\fB}\bu^\infty\cdot(\bn\cdot\tTh_\tU)\d S- \int_{\fV}\btau'_{NN}:\tEh_{\tU}\d V\rb],
\label{velocity_shear}
\end{align}
where $\tF_{ext} = \left[\bF_{ext} \quad 0 \right]^T$. Here, $\bF_{ext}$ is an arbitrary external force acting on the particle. The particle is immersed in a 2D linear flow given by  $\bm{u}^{\infty *} = \bm{A}^{\infty *} \cdot \bm{x}^*$ (dimensionless), in a Cartesian basis we may write
\begin{align}
\bm{A}^{\infty *} = \begin{bmatrix} 1+ \lambda &1 - \lambda&0\\-\left(1- \lambda \right)&-\left(1+ \lambda \right)&0\\0&0&0 \end{bmatrix} 
\end{align} 
where we have scaled length with the radius of the sphere and stresses with $\eta_o \gammad_c$, where $\gammad_c$ is characteristic of the applied strain-rate such that we have $\bm{A}^{\infty*}$ in the form above. It is useful to also decompose $\bu^{\infty*}=\frac{1}{2}\dot{\bgamma}^{\infty*} \cdot \bx^*+\bOmega^{\infty*}\times \bx^*$ into symmetric and antisymmetric parts associated with strain-rate and rotation-rate respectively. Note that $\lambda = -1$ corresponds to purely rotational flow, $\lambda = 0$ is shear flow and $\lambda = 1 $ extensional flow \citep{2Dflow}. 

On evaluating \eqref{velocity_shear} we get the translational velocity of the particle
\begin{align}
\bm{U}^{*} = \lb\{\bI+ \frac{1}{2} Cu^2 \left( 1 - \beta \right) \left( 1- n\right)\left[\frac{942}{2275}\frac{\vert \bm{F}^*_{ext} \vert^2}{(6 \pi)^2}\bI +\frac{695}{539}\vert \dot{\gamma}^{\infty*}\vert^2 \bI + \frac{10037}{7007}\dot{\bgamma}^{\infty*}\cdot\dot{\bgamma}^{\infty*}\right]\rb\}\cdot\frac{\bF^*_{ext}}{6\pi}.
\end{align}
The first two terms on the right hand side correspond to the velocity due to an external force in an otherwise quiescent shear-thinning fluid as in \eqref{mobility}. The remaining terms demonstrates the coupling between the background flow field and external force. When the force is perpendicular to the plane of applied flow, we see that the velocity of the particle is further increased above its quiescent fluid value due to the thinning of the fluid by the external flow field. However, interestingly, for any general direction of the external force, the velocity of the particle may not be in the direction of the forcing. This is due to the lack of symmetry of the background flow field in one direction. It is also worth noting that for a purely rotational flow, one does not see a shear-thinning effect arising from the background flow. 

Using \eqref{velocity_shear} we also evaluate the angular velocity of the sphere, which is given by  
\begin{align}
\bm{\Omega}^* = \bm{\Omega}^{\infty*} + \frac{1}{2}Cu^2 \left( 1- \beta\right) \left(1-n\right) \dfrac{3189}{4004}\left[\frac{\bF^*_{ext}}{6\pi} \times \dot{\bgamma}^{\infty*} \cdot \frac{\bF^*_{ext}}{6\pi} \right].
\end{align}
Note that in the absence of any forcing the sphere rotates with just the background angular velocity just like in a Newtonian fluid where the angular velocity of the sphere in a background flow is independent of viscosity. This was also found in the numerical simulations of  \citet{davino_rotation}. Also, if the external force is along any of the principal directions of strain, or the background flow is purely rotational, the angular velocity of the sphere will be just due to the rotational component of the background flow. However, for an arbitrary direction of the external force, the angular velocity may be different than that imposed by background flow field. 

\section{Suspension of spheres in a shear-thinning fluid}
Suspensions of particles in shear-thinning fluids are encountered in a wide range of chemical, biochemical and material processing industries, and as such there has been considerable interest in studying the flow properties of such suspensions \citep{raju2015, Chhabra_bubble, Gummu_bubble, Kishore_bubble, laven_einstein, Tanner_suspension}. Most of these studies consider particles in power-law fluids, in other words, it is assumed that the strain-rates are large enough so that the fluid rheology is captured by a power law model. Here, we complement these studies by quantifying the first effects of the non-Newtonian rheology of the suspending fluid in the realm of small strain-rates.

We calculate the average stress in a dilute suspension of neutral buoyant rigid spheres in a weakly shear-thinning fluid, subject to a linear background flow
\begin{align}
\bm{u}^{\infty *} = \bm{A}^{\infty*} \cdot \bm{x^*}
\end{align}
as discussed in the previous section. The average stress in a suspension of rigid spheres in a weakly shear-thinning fluid, correct to $\fO\left( Cu^2\right)$, is evaluated as
\begin{equation}
\langle \reallywidehat{\bsigma^*} \rangle= \langle \dot{\bm{\gamma}}^* \rangle   - \frac{1}{2} Cu^2 \left( 1- \beta\right) \left( 1- n \right) \langle  \vert \dot{\gamma}^*\vert^2  \dot{\bm{\gamma}}^* \rangle + \langle \reallywidehat{\bsigma_{p}^* }\rangle,
\label{average_stress}
\end{equation} 
where $\bsigma_p$ is the additional stress within the suspended particles, and the average quantities (denoted with angular brackets) are obtained by taking an ensemble average over all possible configurations of the particles \citep{batchelor_1970, koch_subu, Rallison_2012}. We use the wide hat symbol $\reallywidehat{\phantom{\bsigma^*}}$ to refer to the symmetric and deviatoric component of a second-order tensor and note that  the isotropic terms in the average stress do not contribute to suspension rheology \citep{Einarsson_viscosity, Rallison_2012}.  In a homogeneous and dilute suspension of particles, we know
\begin{equation}
 \langle \reallywidehat{\bsigma_{p}^* }\rangle = n \reallywidehat {\bS^*}
\end{equation}
where $n$ is the particle number density equal to $\phi/ V_p$, where $\phi \ll 1$ is the particle volume fraction, $V_p$ is the volume of a single particle, and $\bS$ is the particle stresslet \citep{batchelor_1970, koch_subu} defined as

\begin{equation}
\bS = \int_{\partial \fB} \dfrac{1}{2} \left[ \bn \cdot \bsigma \br + \br \bn \cdot \bsigma \right] \d S.
\label{stresslet_definition}
\end{equation}

In order to calculate the average stress in the suspension, we start by evaluating the particle stresslet, using the reciprocal theorem  \citep{sebastien_stresslet, elfring17, Einarsson_viscosity}. \citet{elfring17} derives an expression for the stresslet in a weakly non-Newtonian fluid which is (in dimensional form)
\begin{align}
\bm{S} = \bm{S} ^{\infty} -\frac{\eta}{\hat{\eta}} \int_{\partial B} \bm{u}^{\infty} \cdot \left( \bm{n} \cdot \tTh_{E}\right) \d S - \int_{\fV} \btau'_{NN}: \tEh_{E} \d V.
\label{equ_for_stt}
\end{align}
Clearly, evaluation of the stresslet $\bS$ up to $\fO\left( Cu^2 \right)$  using equation \eqref{stresslet_definition} would require calculating the first correction to the flow field in the shear-thinning fluid; however, equation \eqref{equ_for_stt}, obtained using the reciprocal theorem, bypasses this calculation and the first correction to the stresslet is obtained by using the Newtonian flow field.

We first evaluate the stresslet $\bS^{\infty}$ due to the stress $\bsigma^{\infty}$ from the background flow. In the absence of particles, the background stress  $\bsigma^{\infty}$ is simply
  \begin{equation}
 \bsigma^{\infty*} = \bgamma^{\infty*}   -\frac{1}{2}Cu^2\left(1- n\right) \left( 1 - \beta \right) \vert \dot{\gamma}^{\infty*}\vert^2  \dot{\bgamma}^{\infty*} + \: \fO\left( Cu^4 \right), 
 \end{equation}
where $ \dot{\bgamma}^{\infty*} = \bm{A}^{\infty*} + \bm{A}^{\infty*\top}  $ is the applied strain rate. Using the definition of the stresslet \eqref{stresslet_definition}, we therefore have, up to $\fO\left( Cu^2\right)$,
\begin{align}
  \bm{S}^{\infty*} = \dfrac{4\pi}{3}  \dot{\bgamma}^{\infty*}  - \frac{1}{2} Cu^2 \left( 1- \beta \right) \left(1- n \right) \dfrac{4 \pi}{3}  \vert \dot{\gamma}^{\infty*}\vert^2 \dot{\bgamma}^{\infty*}.
  \label{final_stresslet}
\end{align}
Evaluating the integral terms in \eqref{equ_for_stt}, we obtain
 \begin{align}
\bm{S}^* - \bm{S} ^{\infty*} = 2  \pi \dot{\bgamma}^{\infty*} - \frac{1}{2} Cu^2 \left( 1- \beta \right) \left( 1- n\right) \dfrac{68469 \pi }{17017} \vert \dot{\gamma}^{\infty*}\vert^2 \dot{\bgamma}^{\infty*},
\label{disturbance_stresslet}
 \end{align}
and therefore altogether have
 \begin{align}
 \bS^* =  \dfrac{10 \pi}{3} \dot{\bgamma}^{\infty*}  - \frac{1}{2} Cu^2 \left( 1- \beta \right) \left(1- n \right) \left(  \dfrac{273475 \pi}{51051} \right) \vert \dot{\gamma}^{\infty*}\vert^2 \dot{\bgamma}^{\infty*}.
 \label{stresslet_expression}
 \end{align}
The first term on the right hand side is the stresslet in a Newtonian fluid. From the equation above, it can be seen that the total stresslet in a shear-thinning fluid is less than that in Newtonian fluid. 

We now  proceed to calculate the average stress in equation \eqref{average_stress}. We note that the mean Newtonian viscous stress, $\langle \dot{\bm{\gamma}}^* \rangle$, is equal to the bulk applied strain-rate $\dot{\bgamma}^{\infty*}$ \citep{koch_subu, Guazzelli_2009}. The second term on the right hand side in equation  \eqref{average_stress} can be evaluated by first performing a formal ensemble average  based on the ergodic hypothesis \citep{Rallison_2012} on $\langle  \vert \dot{\gamma}^*\vert^2  \dot{\bm{\gamma}}^* \rangle$. Writing the strain-rate in terms of the mean and fluctuating components $\dot{\gamma}^*_{ij}=\langle\dot{\gamma}^*_{ij}\rangle + \dot{\gamma}_{ij}^{'*}$, we obtain
\begin{align}
\langle{ \vert\dot{\gamma}^*\vert^{2}} \rangle &=  \langle{\dot{\gamma}^*_{ij} \dot{\gamma}^*_{ij}}\rangle   =  \langle\dot{\gamma}^*_{ij}\rangle\langle\dot{\gamma}^*_{ij}\rangle + \langle\dot{\gamma}_{ij}^{'*} \dot{\gamma}_{ij}^{'*}\rangle, 
 \end{align}
where dashed quantities are fluctuating values. Here, we have used the fact $\left<\gammad_{ij}^{'} \langle\gammad_{ij}\rangle\right> = 0$ and as such, obtain $ \vert\dot{\gamma}^*\vert^{2'} = 2  \dot{\gamma}_{ij}^{'*} \langle\dot{\gamma}^*_{ij}\rangle + \dot{\gamma}_{ij}^{'*} \dot{\gamma}_{ij}^{'*} - \langle\dot{\gamma}_{ij}^{'*} \dot{\gamma}_{ij}^{'*}\rangle $. Using these, the ensemble average: 
 \begin{align}
\langle{ \vert\dot {\gamma}\vert ^{2*}  \bm{\dot{\gamma}}^*} \rangle =  \langle \vert\dot {\gamma}\vert ^{2*}\rangle \: \langle\bm{\dot{\gamma}}^*\rangle + \langle\vert\dot {\gamma}\vert ^{2'*}  \bm{\dot{\gamma}}^{'*}\rangle
= \left( \langle{\dot{\gamma}_{ij}^*}\rangle \:  \langle{\dot{\gamma}^*_{ij}}\rangle + \langle\dot{\gamma}_{ij}^{'*} \dot{\gamma}_{ij}^{'*}\rangle\right) \langle{\bm{\dot{\gamma}}^*}\rangle  + 2 \langle \dot{\gamma}_{ij}^{'*} \langle{\dot{\gamma}^*_{ij}}\rangle  \bm{\dot{\gamma}}^{'*}\rangle + \langle{\dot{\gamma}_{ij}^{'*} \dot{\gamma}_{ij}^{'*}  \bm{\dot{\gamma}}^{'*}}\rangle.
\label{terms_to_be_ag}
 \end{align}
 
Performing this ensemble averaging step has been shown  \cite{Rallison_2012} to remove terms that in a volume average give rise to divergent integrals for dilute suspensions in second-order fluids \citep{greco_rheology, Housiadas_2009}. We now replace ensemble average by volume average in \eqref{terms_to_be_ag}, and evaluate the quantities both inside the solid spheres and in the fluid volume, noting that inside the solid particles $\dot{\bgamma}^{'*} = - \dot{\bgamma}^{\infty*}$, since the total strain-rate inside the particle is zero \citep{koch_subu}.

 Following \citet{koch_subu}, evaluation of \eqref{terms_to_be_ag} gives
 \begin{align}
 \langle \vert \dot{\gamma}^*_0\vert^2 \dot{\bm{\gamma}}^*_0 \rangle = \left( 1+ \dfrac{125 \phi}{28}\right)  \vert \dot{\gamma}^{\infty*}\vert^2 \dot{\bgamma}^{\infty*}.
 \label{value_of_stress}
 \end{align}

Summing all the contributions to the average stress in equation \eqref{average_stress}, we have, finally,
 \begin{align}
  \langle \widehat{\bm{\sigma}^*} \rangle = \left[ 1 + 2.5 \phi   -\frac{1}{2}Cu^2 \left( 1- \beta \right) \left(1 - n \right)  \left( 1+ b \phi \right) \vert \dot{\gamma}^{\infty*}\vert^2 \right] \dot{\bgamma}^{\infty*}.
 \end{align}
where $b = 288675/34034$. The term inside the square bracket gives the effective viscosity of the suspension. The presence of particles thickens the fluid at the leading order (Einstein viscosity) where as at $\fO\left(Cu^2\right)$ it decreases the effective viscosity due to enhanced thinning of the fluid. We also note that decreasing $n$ linearly reduces the total correction to fluid viscosity from the Einstein correction as discussed by \citet{Tanner_suspension} for dilute suspensions in power law fluids.  The presence of particles in a shear-thinning fluid could lead to interesting rheological behaviour when the thickening and thinning effects of particles compete at the same order. Also, the form of above expression suggests that a dilute suspension of rigid spheres in a Carreau fluid will behave as a Carreau fluid. In fact, our results agree with recent results by \citet{Domurath_2015} who use a numerical homogenization technique to obtain the effective viscosity of a dilute suspension in a Bird-Carreau model and find that the effective viscosity too can be modelled using a Bird-Carreau model with modified values of the parameters.

\section{Conclusion}
In this work, we considered a few problems involving spheres in shear-thinning fluids at zero Reynolds number. Using the reciprocal theorem, we analytically demonstrated how shear-thinning rheology may lead to qualitative changes in the particle dynamics compared to Newtonian fluids. Specifically, we showed that the translational and rotational dynamics of a sphere are coupled in shear-thinning fluids which can lead to interesting dynamics in problems involving sedimentation of rotating spheres (a setup which may be used as a rheometer) and sedimentation in a background flow field. We also showed that for two equal spheres sedimenting along the line joining their centres, the symmetry arguments used in Newtonian fluids will predict the observed result in a generalised Newtonian fluid.   Although these two spheres will sediment maintaining their initial distance of separation, the variation of the shear-thinning effects with initial separation distance is non-monotonic. Finally, we considered a dilute suspension of spheres in a weakly shear-thinning fluid and showed that the resulting suspension will also be a weakly shear-thinning fluid with a viscosity that varies due to competing effects arising from the presence of particles: the particles thicken the fluid (the Einstein viscosity correction) but also increase effective strain-rates thereby enhancing shear-thinning. At higher strain-rates, outside the scope of our weakly non-linear assumption, it would be interesting to investigate strain rates at which the thinning effect supersedes the thickening one.

\acknowledgements
Funding from the Natural Sciences and Engineering Research Council of Canada (NSERC) is gratefully acknowledged.

\section{Appendix}

Here we present some expressions which were used to evaluate the integrals. Additional details on related expressions for both passive and active particles can be found in the recent work of \citet{nasouri2018note}.  

For a single sphere of radius $a$ in a Newtonian fluid with viscosity $\etah$ 
\begin{align}
\bRh_{FU} &= 6 \pi \etah a \bI,\\
 \bRh_{L\Omega} &= 8 \pi \etah a^3 \bI, \\
  \bRh_{LU} &= 0,\\
    \bRh_{F\Omega} &= 0.
\end{align}
Additionally, 
\begin{equation}
\bn \cdot \tTh_\tU = - \dfrac{3\etah}{2a} \left[ \bI  \quad 2\bTheta \right],
\end{equation}
where $\Theta_{ij} = \epsilon_{ijk} x_k$ and $\bn$ is the unit normal to the surface. 

The entities corresponding to $\hat{\dot{\bgamma}}/2 = \tEh_{\tU} \cdot \tUh$
\begin{align}
[\tEh_{U}]_{ijk} &= \dfrac{3a x_k}{4 r^3} \left(\delta_{ij} - \dfrac{3 x_i x_j}{r^2} \right) + \dfrac{3a^3}{4r^5} \left[ x_k \left( -\delta_{ij} + \dfrac{5 x_i x_j}{r^2}\right) - x_i \delta_{jk} - x_j \delta_{ik}\right] \\
[\tEh_\Omega]_{ijk} &= - \dfrac{3 a^3}{2r^5} \left(x_i x_l \epsilon_{ljk} + x_j x_l \epsilon_{lik}\right). 
\end{align}
are detailed in \citep{PRFStone}. 
Relevant to the stresslet calculation, we have 
\begin{equation}
\begin{split}
[\tEh_E]_{klij} = & \dfrac{\etah a^5}{2r^5} \left( \delta_{ik} \delta_{jl} +  \delta_{jk} \delta_{il}  \right) + \dfrac{5 \etah a^3} {4 r^5} \left(  \delta_{il} x_j x_k  + \delta_{ki} x_j x_l + \delta_{jl} x_i x_k  + \delta_{kj} x_i x_l  \right)\\ &- \dfrac{5 \etah a^5}{2r^7} \left( \delta_{kl} x_i x_j + \delta_{jl} x_i x_k  + \delta_{il} x_j x_k  + \delta_{jk} x_i x_l + \delta_{ik} x_j x_l  \right) + 5 \etah \left( \dfrac{7 a^5}{2r^9} - \dfrac{5 a^3}{2r^7}\right) x_i x_j x_k x_l + \dfrac{5 \etah a^3 }{2r^5} \delta_{kl} x_i x_j 
\end{split}
\end{equation}
\begin{equation}
\begin{split}
[\tTh_E]_{klij} = & \dfrac{\etah a^5}{r^5} \left( \delta_{ik} \delta_{jl} +  \delta_{jk} \delta_{il}  \right) + \dfrac{5 \etah a^3} {2 r^5} \left(  \delta_{il} x_j x_k  + \delta_{ki} x_j x_l + \delta_{jl} x_i x_k  + \delta_{kj} x_i x_l  \right)\\ &- \dfrac{5 \etah a^5}{r^7} \left( \delta_{kl} x_i x_j + \delta_{jl} x_i x_k  + \delta_{il} x_j x_k  + \delta_{jk} x_i x_l + \delta_{ik} x_j x_l  \right) + 5 \etah \left( \dfrac{7 a^5}{r^9} - \dfrac{5 a^3}{r^7}\right) x_i x_j x_k x_l 
\end{split}
\end{equation}
which can be found in the supplementary information of \citep{sebastien_stresslet}.

For the problem involving two spheres sedimenting along their common axis, the stream functions for the two auxiliary cases in Newtonian fluids, two spheres moving with the same velocity and two spheres approaching each other with equal speed, were reported by \citet{stimson1926motion} and \citet{BRENNER1961242}, respectively. The stream functions are expressed in the form of infinite series solutions.  To ensure convergence, we considered around the first 30 to 40 terms of the series. The stream functions are used to calculate the strain-rate tensors corresponding to $\tEh_{\overline{U}}$ (Stimson-Jeffery) and $\tEh_{\Delta U}$ (Brenner, Maude). The stress tensor $\btau_{NN}$ for the case considered in the manuscript is evaluated using strain-rates from the Stimson-Jeffery solution.  

For the limiting case of two spheres touching each other and sedimenting, we use the solution of the problem in Newtonian fluids by \citet{cooley_1969} to evaluate both $\btau_{NN}$ and $\tEh_{U}$. The stream function for this case is expressed in form of a definite integral from zero to infinity (equation 3.4 in the reference). A non-infinite value of the upper limit of the integral has to be chosen for evaluation; we find that convergence of the solution is achieved at a value of around 15.

\bibliography{reference}

\end{document}